\documentstyle{mn}
\input{epsf}
\newcommand{\be}{\begin{eqnarray}}
\newcommand{\ee}{\end{eqnarray}}

\title[Large Scale Structure in the SDSS Galaxy Survey]
{Large Scale Structure in the SDSS Galaxy Survey}
\author[Doroshkevich, Tucker \& Allam]
       {A. Doroshkevich$^{1,2}$ D.L. Tucker$^{3}$ \& S. Allam$^{3,4}$\\
	$1$Theoretical Astrophysics Center,
          Juliane Maries Vej 30,
          DK-2100 Copenhagen \O, Denmark\\
	$2$Keldysh Institute of Applied Mathematics,
                        Russian Academy of Sciences,
                        125047 Moscow,  Russia\\
	$3$Fermi National Accelerator Laboratory, MS 127, P.O. Box 500, 
	Batavia,\\ IL 60510 USA \\
	$4$National Research Institute for Astronomy \& Geophysics, Helwan\\ 
	Observatory, Cairo, Egypt
}

\begin{document}
\maketitle

\begin{abstract}
The Large Scale Structure (LSS) in the galaxy distribution is
investigated using the Sloan Digital Sky Survey Early Data Release
(SDSS EDR). Using the Minimal Spanning Tree technique we have
extracted sets of filaments, of wall-like structures, of galaxy
groups, and of rich clusters from this unique sample.  The physical
properties of these structures were then measured and compared with
the expectations from Zel'dovich' theory.

The measured characteristics of galaxy walls were found to be
consistent with those for a spatially flat $\Lambda$CDM
cosmological model with $\Omega_m\approx$ 0.3 and $\Omega_\Lambda
\approx$ 0.7, and for Gaussian initial perturbations with a 
Harrison -- Zel'dovich power spectrum. Furthermore, we found that 
the mass functions of groups and of unrelaxed structure elements 
generally fit well with the expectations from Zel'dovich' theory, 
although there was some discrepancy for lower mass groups which 
may be due to incompleteness in the selected sample of groups.
We also note that both groups and rich clusters tend to prefer the
environments of walls, which tend to be of higher density, rather 
than the environments of filaments, which tend to be of lower density.

Finally, we note evidence of systematic differences in the properties
of the LSS between the Northern Galactic Cap stripe and the Southern 
Galactic Cap stripe -- in particular, in the physical properties of 
the walls, their spatial distribution, and the relative numbers of 
clusters embedded in walls.  Because the mean separation of walls 
is $\approx$ 60 -- 70$h^{-1}$~Mpc, each stripe only intersects a few 
tens of walls. Thus, small number statistics and cosmic variance are 
the likely drivers of these systematic differences.

\end{abstract}


\section{Introduction}

With the advent of the Durham/UKST Galaxy Redshift Survey (DURS,
Ratcliffe et al. 1996) and the Las Campanas Redshift Survey (LCRS,
Shectman et al. 1996), the galaxy distribution on scales up to
$\sim$300 $h^{-1}$~Mpc could be studied. Now these investigations can
be extended using the public data sets from the Sloan Digital Sky
Survey Early Data Release (SDSS EDR; Stoughton et al. 2002), which
contains $\approx$ 30\,000 galaxies in two slices for distances $D\leq
600 h^{-1}$Mpc.

The analysis of the spatial galaxy distribution in the DURS and the
LCRS has revealed that the Large Scale Structure (LSS) is composed of
walls and filaments, that galaxies are divided roughly equally into
each of these two populations (with few or no truly isolated
galaxies), and that richer walls are linked to the joint random
network of the cosmic web by systems of filaments (Doroshkevich et al.
2000; 2001).  Furthermore, these findings are consistent with results
obtained for simulations dark matter (DM) distributions (see, e.g.,
Cole et al. 1998; Jenkins et al. 1998) and for mock galaxy catalogues
based upon DM simulations (Cole et al. 1998).

The quantitative statistical description of the LSS is in itself an
important problem.  Beyond that, though, the analysis of rich
catalogues can also provide estimates for certain cosmological
parameters and for the characteristics of the initial power spectrum
of perturbations. To do so, some theoretical models of structure
formation can be used.

The close connection between the LSS and Zel'dovich' pancakes has been
discussed by Thompson \& Gregory (1978) and by Oort (1983).  Now this
connection is verified by the comparison of the statistical
characteristics of observed and simulated walls with theoretical
expectations (Demia\'nski \& Doroshkevich 1999; 2002, hereafter DD99
\& DD02) based on the Zel'dovich theory of nonlinear gravitational
instability (Zel'dovich 1970; Shandarin \& Zel'dovich 1989). This
approach connects the characteristics of the LSS with the main
parameters of the underlying cosmological scenario and the initial
power spectrum, and it permits the estimation of some of these
parameters using the measured properties of walls. It was examined
with the simulated DM distribution (DD99; Demia\'nski et al. 2000),
and was found that, for sufficiently representative samples of walls,
a precision of better than 20\% can be reached.
 
Effective methods of the statistical description of the LSS were
developed by Demia\'nski et al. (2000) and Doroshkevich et al.  (2000;
2001), who applied them to DM simulations and to the DURS and the
LCRS. In this paper we apply the same approach to the SDSS EDR, a
sample from which we can obtain more representative and more precise
measures of the properties of the LSS and the initial power spectrum
of perturbations.

We widely use the Minimal Spanning Tree technique.  With this
technique, we can quantitatively describe the sample under
investigation, divide the sample into physically motivated subsamples,
and extract different sets of the LSS elements.  This technique allows
us to discriminate between filamentary and wall--like structure
elements located presumably within low and high density regions and to
estimate their parameters for the different threshold overdensities
bounding them. The same technique allows us to extract sets of groups
and rich clusters of galaxies and to measure some of their properties.

Comparison of the observed characteristics of walls with the
theoretical expectations (DD99; DD02) demonstrates that the observed
galaxy distribution is consistent with Gaussianity initial
perturbations and that the walls are the recently formed, partly
relaxed Zel'dovich'pancakes. The mean basic characteristics of the
walls are consistent with those theoretically expected for the initial
power spectrum measured by the CMB observations.

In this paper we also analyse the mass functions of structure elements
selected for a variety of boundary threshold overdensities.  We show
that these functions are quite similar to the expectations of the
Zel'dovich' theory, which generalizes the Press -- Shechter formalism
for any structure elements.  In addition, the theory indicates that the
interaction of large and small scale perturbations can be important
for the formation of the observed LSS mass functions. Our analysis
demonstrates that this interaction is actually seen in the influence
of environment on the characteristics of groups of galaxies.

This paper is organized as follows: In Secs. 2 we describe the sample
of galaxies which we extracted from the SDSS EDR and the method we
have employed to correct for radial selection effects.  In Sec. 3 we
establish the general characteristics of the LSS. More detailed
descriptions of filamentary network and walls can be found in Secs. 4
and 5, respectively. In Secs. 6 and 7 we discusse the probable
selected clusters of galaxies and the mass function of structure
elements. We conclude with Sec.  8 where a summary and a short
discussion of main results are presented.

\section{The SDSS Early Data Release}

We use as our observational sample the SDSS EDR (Stoughton et
al. 2002), which is the first public release of data from the SDSS
(Fukugita et al. 1996, Gunn et al.\ 1998, and York et al. 2000).  

The imaging data for the SDSS EDR encompasses 462~sq~deg of sky -- a
$2.5^{\circ} \times 90^{\circ}$ equatorial slice in the North Galactic
Cap (Runs 752 \& 756), a $2.5^{\circ} \times 66^{\circ}$ equatorial
slice in the South Galactic Cap (Runs 94 \& 125), and about 68~sq~deg
in the direction of the SIRTF First Look Survey (Runs 1336, 1339,
1356, \& 1359).  The EDR also contains followup spectra, which are
available for all but two $2.5\times2.5$ fields in the North Galactic
Cap slice.

We obtained our SDSS EDR sample via the SDSS Query Tool ({\tt
sdssQT})\footnote{ http://archive.stsci.edu/sdss/software/\#sdssQT}, a
standalone interface to the SDSS Catalog Archive Server.  The exact 
query used in documented in the Appendix.

In our analysis here, we ignore the SIRTF fields and consider just the
two equatorial slices.  In particular, we consider four samples
based upon these two slices:  
\begin{itemize}
\item N-600, the northern sample for $D \leq 600 h^{-1}$Mpc (16\,883 galaxies)
\item S-600, the southern sample for $D \leq 600 h^{-1}$Mpc (12\,428 galaxies)
\item N-380, the northern sample for $D \leq 380 h^{-1}$Mpc (13\,698 galaxies)
\item S-380, the southern sample for $D \leq 380 h^{-1}$Mpc  (9\,924 galaxies)
\end{itemize}

\subsection{Correction for the radial selection effects}

In Figure 1 we plot the radial distributions of galaxies in the N-600
and S-600 samples.  Note that the radial selection effects clearly
seen in these two distributions are quite successfully fit by curves
describing a selection function of the form
\be
f_{gal}(D)\propto D^2\exp[-(D/R_{sel})^{3/2}],\quad R_{sel}\approx 
190 h^{-1}{\rm Mpc}\,,
\label{sel}
\ee
where $D$ is a galaxy's radial distance and $R_{sel}$ is the 
selection scale (Baugh \& Efstathiou 1993). 

\begin{figure}
\centering
\epsfxsize=7.cm
\epsfbox{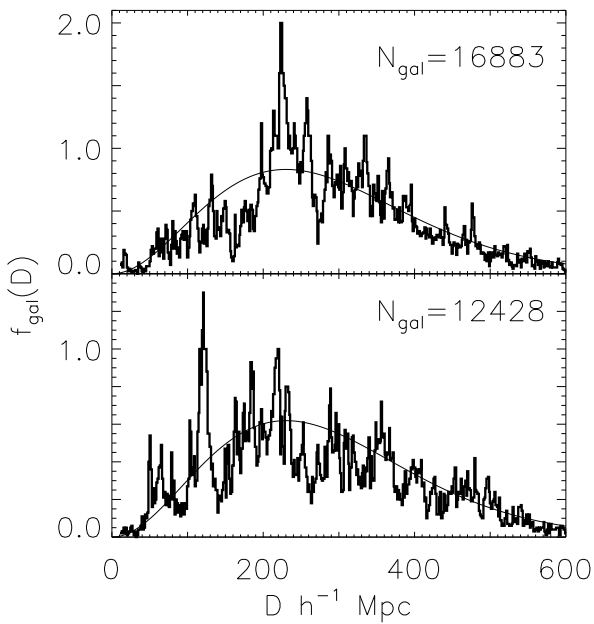}
\vspace{1.4cm}
\caption{The radial galaxies distributions in the samples 
N-600 (top panel) and S-600 (bottom panel). The selection 
function (\ref{sel}) is plotted by solid lines.
} 
\end{figure}

In some applications, like when we want to correct a measure of the
observed density to a measure of the true density, we would like to
use equation (\ref{sel}) to correct for the radial selection effects
after the fact.  An example of such a case is calculating a group's 
or cluster's true richness based upon the observed number of galaxies 
it contains (Sec. 6 \& 7).

In other applications, however, like in searching for groups or
clusters in a magnitude-limited sample, we want to make a preemptive 
correction for the radial selection effects.  For example, in a standard 
friends-of-friends percolation algorithm (e.g., Huchra \& Geller 1982), 
this is done by adjusting the linking length as a function of radial 
distance. Here, instead, we employ the rather novel approach of 
adjusting the radial distances themselves; so, instead of the measured 
radial distance, we use a modified radial distance, $D_{md}$, where 
\be
D_{md}^3 = 2R_{sel}^3(1-[1+(D/R_{sel})^{3/2}]\exp[-(D/R_{sel}
)^{3/2}])\,.
\label{ra}
\ee 
The radial variations of the normalized number density of galaxies for
both samples from Figure~1 are plotted in Figure~2. As is seen from
this figure, the modified radial distances for the galaxies suppresses
the very large-scale trends and emphasizes the smaller scale random 
variations in the density.

This correction is more important for the more distant regions of our
samples ($D\geq 350 h^{-1}$Mpc), which contain only $\sim$20\% of
galaxies. Thus, in the following analyses, we apply this correction
only to the two deeper samples, the N-600 and the S-600.  Of course,
it cannot restore the lost information about the galaxy distribution
in these regions, but it does help compensate the strong drop in the
observed galaxy density at these distances and to apply the standard
methods of investigation for the full catalogues with the depth
600$h^{-1}$Mpc.

\begin{figure}
\centering
\epsfxsize=7.cm
\epsfbox{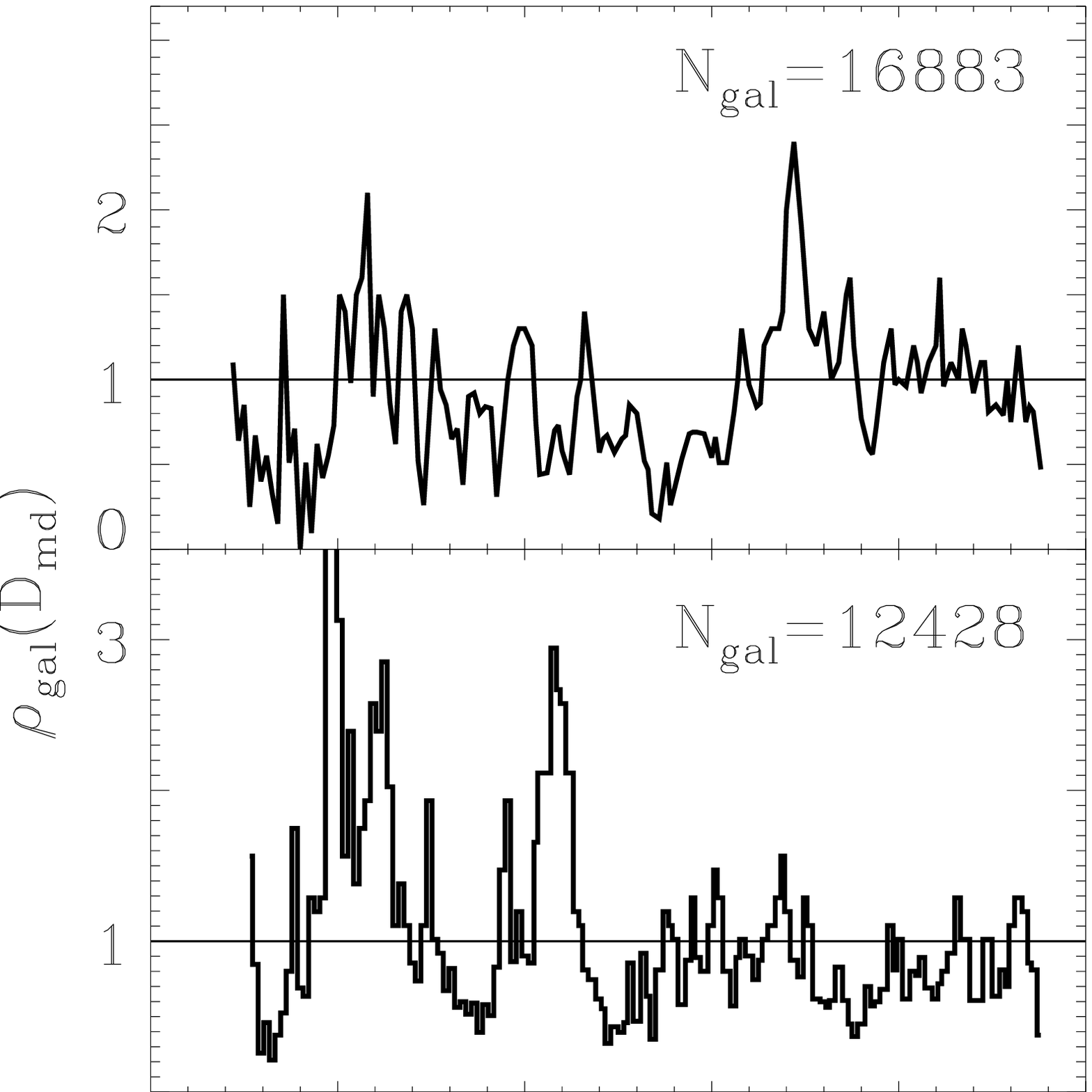}
\vspace{1.4cm}
\caption{The normalized mean galaxy density in the modified 
samples N-600 (top panel) and S-600 (bottom panel). 
} 
\end{figure}

\section{General characteristics of observed large scale structure}

To characterize the general properties of large scale spatial galaxy 
distribution we use the Minimal Spanning Tree (MST) technique applied 
to both directly observed samples of galaxies and to samples corrected 
for the selection effect. The MST technique was first discussed by 
Barrow, Bhavsar \& Sonoda (1985) and by van de Weigaert (1991). Its 
applications for the quantitative description of observed and 
simulated catalogues of galaxies were discussed in Demia\'nski et al. 
(2000) and Doroshkevich et al. (2000, 2001). 

\subsection{Wall-like and filamentary structure elements}

With the MST technique we can demonstrate that the majority of 
galaxies is concentrated within wall-like structures and filaments 
which connect walls to joint random network of the cosmic web. The 
internal structure of both walls and filaments is complex. Thus, 
wall-like structures incorporate some fraction of filaments and both 
walls and filaments incorporate high density galaxy clouds. In 
particular, clusters of galaxies are usually situated within richer 
walls. 

As is well known, for larger galaxy separations a Poisson-like point
distribution can be expected for galaxies within structures.  As a
result, the probability distribution function of MST edge lengths (PDF
MST), $W_{MST}(l)$, characterizes the geometry of the galaxy
distribution.  For the 1D and 2D Poissonian distributions typical for
filaments and walls, respectively, $W_{MST}(l)$ is described by
exponential and Rayleigh functions, namely,
\begin{equation}
W_{MST}(l)=\langle l\rangle^{-1}\exp(-l/\langle l\rangle)\,,
\label{W} 
\end{equation}
\[
W_{MST}(l)=2l/\langle l^2\rangle\exp(-l^2/\langle l^2\rangle)\,. 
\]
Comparison of measured and expected PDFs MST allows us to demonstrate
the existence of these two types of structure elements and to make
approximate estimates of their richness.

In Fig.~3, we see plotted the $W_{MST}(l_{MST})$'s for the N-380 and 
S-380 samples. Note that these $W_{MST}(l_{MST})$'s are well fitted 
to a superposition of Rayleigh (at $l_{MST}\leq \langle l_{MST}\rangle$) 
and exponential (at $l_{MST}\geq \langle l_{MST}\rangle$) functions, 
thus confirming the high degree of galaxy concentration within the  
population of high density rich wall-like structures and less rich 
filaments. 

\begin{figure}
\centering
\epsfxsize=7.cm
\epsfbox{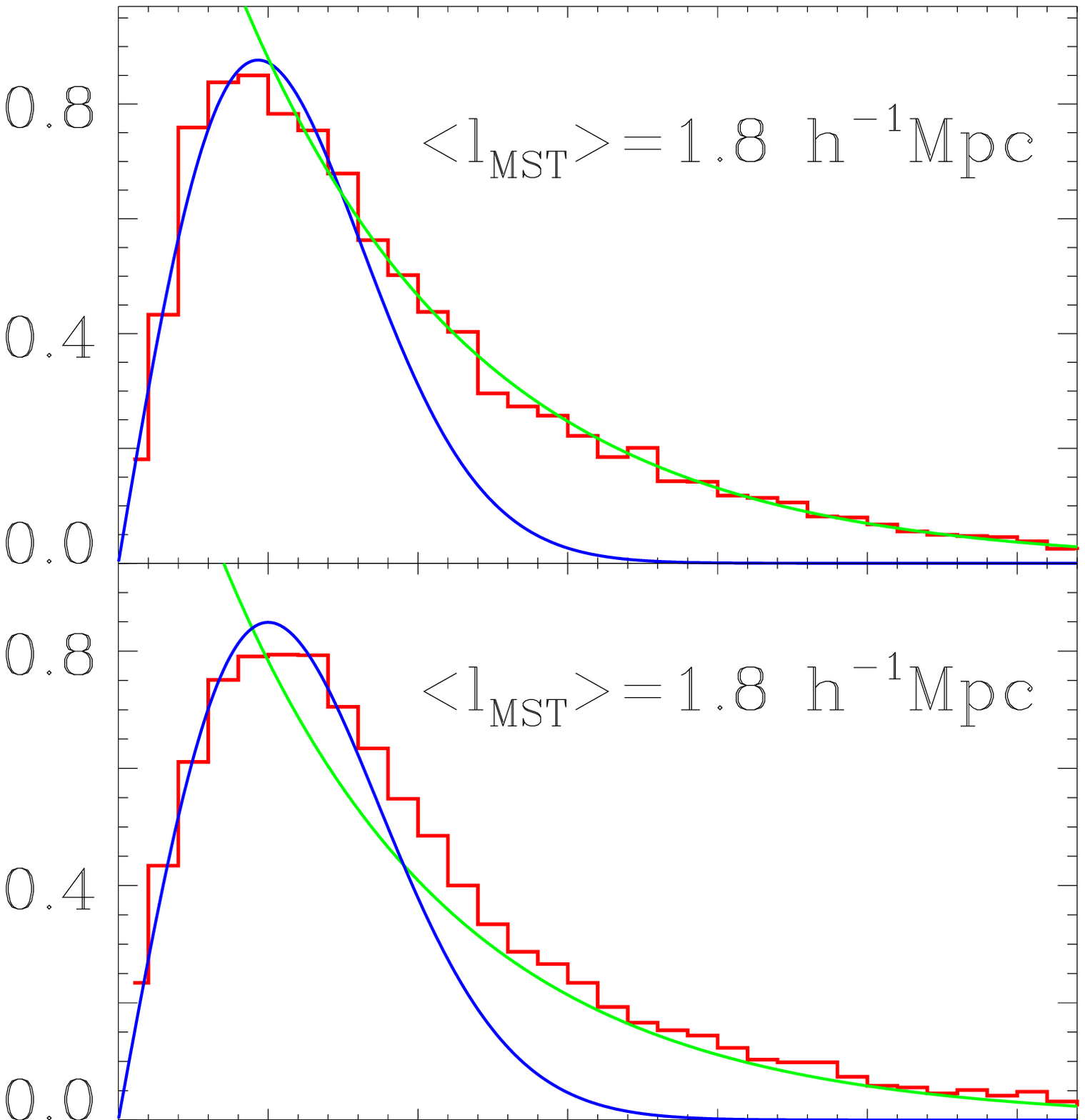}
\vspace{1.cm}
\caption{Distribution functions of MST edge lengths in redshift 
space for samples N-380 (top panel) and S-380 (bottom panel) are 
plotted by red lines. Rayleigh ($\approx 70\%$ of galaxies) and 
exponential fits are plotted by blue and green lines.
} 
\end{figure}

\subsection{High and low density regions}

The methods for dividing a sample into subsamples of wall--like
structures and filaments were proposed and tested in our previous
publications (Demia\'nski et al. 2000; Doroshkevich et al. 2000,
2001).  The first step is to make a rough discrimination between
the high and low density regions (HDRs and LDRs).

Such discrimination can be easy performed for a given overdensity
contour bounding the clusters and a given threshold richness of
individual elements. In particular, to characterize the overdensity,
$\delta_{thr}$, we can use the threshold linking length, $r_{lnk}$,
and a relation familiar from friends-of-friends algorithms (Huchra \& 
Geller 1982):
\be 
\delta_{thr} = 3/[4\pi\langle n_{gal}\rangle r_{lnk}^3]\,.
\label{rlnk}
\ee

In both the N-380 and S-380 samples, wall-like high density regions
(HDRs) were identified with clusters found for a threshold richness
$N_{thr}=$ 50 and a threshold overdensity contour bounding the cluster
equal to the mean density, $\delta_{thr}=1$.  For comparison, other
samples of HDRs were separated with the same $N_{thr}=$ 50 but a
smaller threshold overdensity contour of $\delta_{thr}=0.75$. These
samples of HDRs contain 45\% and 51\% of all galaxies, respectively.
The samples of low density regions (LDRs), which are occupied by
filaments and poor groups of galaxies, are complementary to the HDRs
in that the LDRs are simply the leftovers from the original total 
samples after the HDRs have been removed.

\begin{figure}
\centering
\epsfxsize=7.cm
\epsfbox{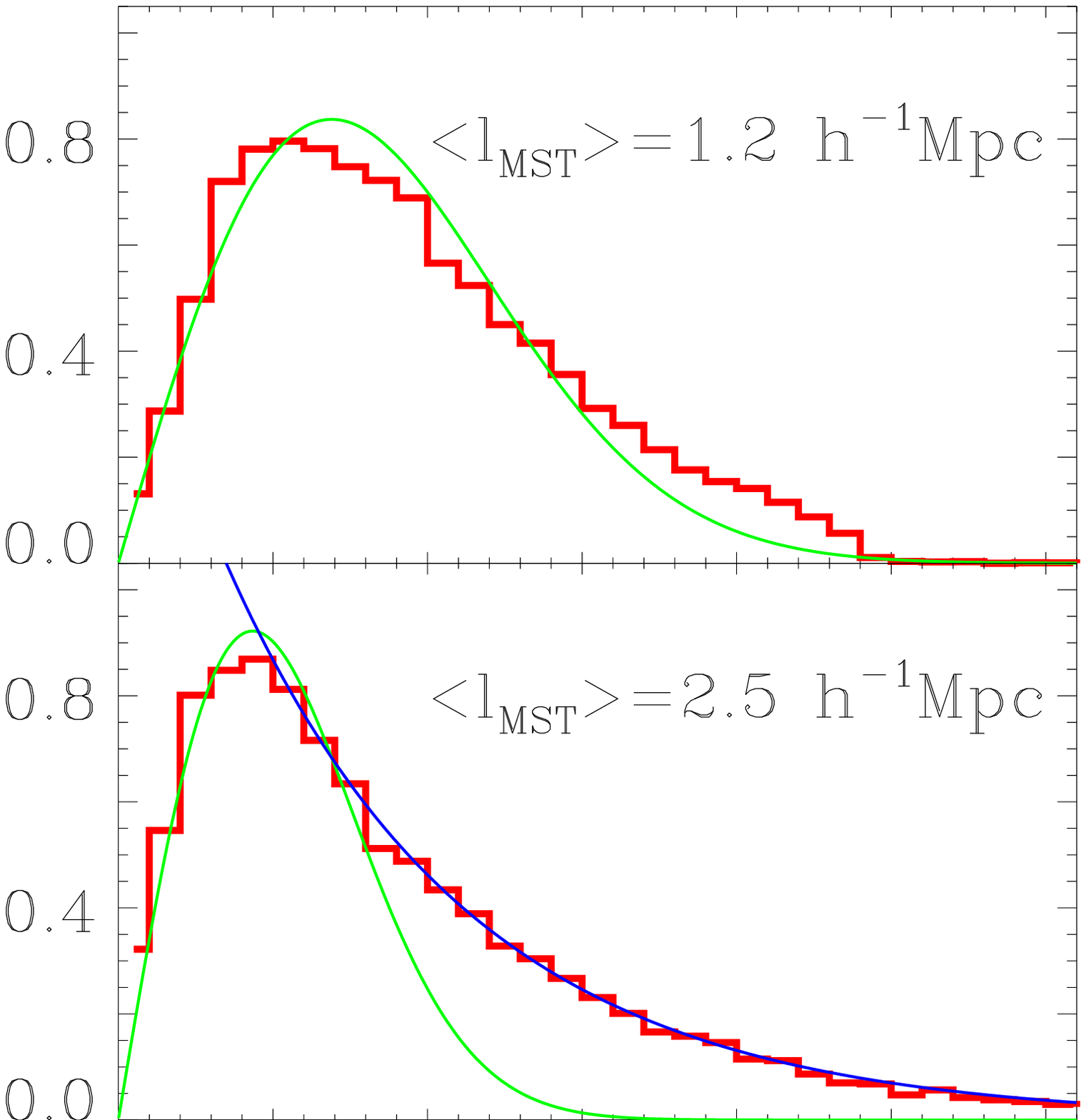}
\vspace{1.3cm}
\caption{Distribution functions of MST edge lengths in redshift 
space for the HDRs (top panel) and LDRs (bottom panel). Rayleigh 
and exponential fits are plotted by green and blue lines.
} 
\end{figure}

In Figure~4, the $W_{MST}(l)$ plotted for the HDRs is very similar to
a Rayleigh function, thus confirming the sheet-like nature of the
observed galaxy distribution within the HDRs. 

The $W_{MST}(l)$ plotted for the LDRs also fits well to a Rayleigh
function -- at least for small edge lengths -- indicating that 
$\sim$ 60\% of LDR galaxies are concentrated within less
massive elliptical and sheet-like clouds.  The LDR $W_{MST}(l)$
for larger edge lengths, however, appears to be closer to an
exponential function, indicating that the remaining $\sim$ 40\%
of LDR galaxies and some part of clouds are in filamentary structures.

The mean edge lengths, $\langle l_{MST}\rangle$, found for HDRs and
LDRs separately, differ by about a factor of two, indicating that the
difference in the mean density of HDRs and LDRs is about an order of
magnitude.

\subsection{Morphology of the structure elements}

With the MST technique we can extract from within HDRs and LDRs
themselves subsamples of structure elements for various threshold
overdensities.  We can then suitably characterize the morphology of
each structure element by comparing the sum all edge lengths within
its full tree, $L_{sum}$, with the sum of all edge lengths within the
tree's trunk, $L_{tr}$, which is the longest path that can be traced
along the tree without re-tracing any steps:
\be
\epsilon =L_{tr}/L_{sum}\,.
\label{trs}
\ee

For filaments, we can expect that the lengths of the full tree 
and of the trunk are similar to each other, whereas for clouds and walls 
these lengths are certainly very different. This approach takes 
into account the internal structure of each element rather than 
the shape of the isodensity contour bounding it, and, in this 
respect, it is complementary to the Minkowski Functional technique 
(see, e.g., Schmalzing at al. 1999). 

\begin{figure}
\centering
\epsfxsize=7.cm
\epsfbox{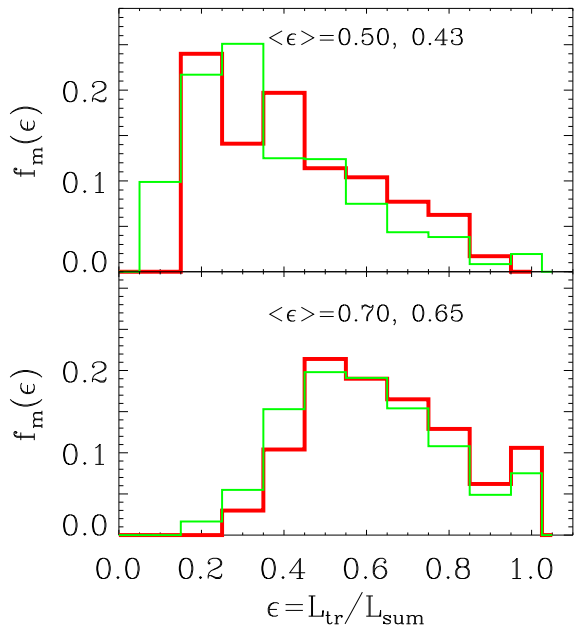}
\vspace{1.3cm}
\caption{Mass functions of structure elements, $f_m(\epsilon)$, 
$\epsilon=L_{tr}/L_{sum}$ for the structure elements selected 
within HDRs with linking lengths $r_{link}=2. ~\&~2.4 h^{-1}$Mpc 
(top panel, red and green lines) and within LDRs with linking 
lengths $r_{link}=3.2 ~\&~3.6 h^{-1}$Mpc (bottom panel, red and 
green lines). 
} 
\end{figure}

However, even this method cannot discriminate between the wall-like
and elliptical clouds and those rich filaments having many long branches
for which again $\epsilon\leq$ 1. This means that both the PDF of this
ratio, $W(\epsilon)$, and the corresponding mass function,
$f_m(\epsilon)$, are continuous functions and the morphology of
structure elements can be more suitably characterized by the degree of
filamentarity and `wall-ness'.  This also means that we can only 
hope to distinguish statistical differences between the morphologies of
structure elements in HDRs and the morphologies of structure elements in
LDRs.

The selection of clusters within HDRs and LDRs was performed for 
two threshold linking lengths, $r_{lnk}=2. ~\&~2.4 h^{-1}$Mpc for 
HDRs, and $r_{lnk}=3.2 ~\&~3.6 h^{-1}$Mpc for LDRs. The distribution 
functions of the ratio, $W(\epsilon)$, are found to be close to Gaussian 
with $\langle \epsilon\rangle\approx 0.5 ~\&~ 0.70$ for HDRs and 
LDRs, respectively. The mass functions, $f_m(\epsilon)$, plotted 
in Fig. 5 for the same linking lengths, are shifted to left (for 
HDRs) and to right (for LDRs) in respect to the middle point.  

These results verify the objective nature of the differences in
the structure morphologies in HDRs and LDRs. 

\section{Typical size of the filamentary network}

What is typical scale of the network of filaments spanning the LDRs?
To estimate a measure of the cell size of the filamentary network, 
we extract filaments from the LDRs using clustering analysis like 
in Sec.~3.2.  We then measure the distance between branch points 
along the trunk of these filamentary clusters. We take as the cell 
size of the filamentary network the mean distance between branch 
points averaged over all filaments. This definition of the filamentary 
network cell size differs from our definition in previous papers 
(e.g., Doroshkevich et al. 1996), where this cell size was defined 
as the mean free path between filaments. The present definition tends
to yield cell sizes that are typically a factor of 1.5 -- 2 smaller
than those yielded by the mean-free-path definition. 

Filaments can be extracted from the LDRs using different threshold
overdensities (different linking lengths).  However, filaments are
connected to the network only for larger linking lengths; thus, the
typical measured cell size depends upon the thresholds used. The
distribution function of the separation of the branch points, 
$W(l_{br})$, is plotted in Fig. 6 for two linking lengths, $r_{lnk}
=3.2 ~\&~3.6 h^{-1}$Mpc, which correspond to the threshold 
overdensities $\delta_{thr}=$ 0.66 \& 0.47.  This distribution 
function is roughly fitted by expression 
\be
W(l_{br})dl_{br}\approx 42 x^{2.5}\exp(-4.1 x)~dl_{br},
\quad x=l_{br}/\langle l_{br}\rangle\,,
\label{lbr}
\ee
\[
\langle l_{br}\rangle\approx 9.5~\&~ 11 h^{-1}{\rm Mpc}\,.
\]

These results are consistent with those obtained in Doroshkevich et
al. (1996; 2001), where the mean free path between filaments for 
the Las Campanas Redshift Survey was found to be $\sim 12 - 
17 h^{-1}$Mpc.

\begin{figure}
\centering
\vspace{0.2cm}
\epsfxsize=7cm
\epsfbox{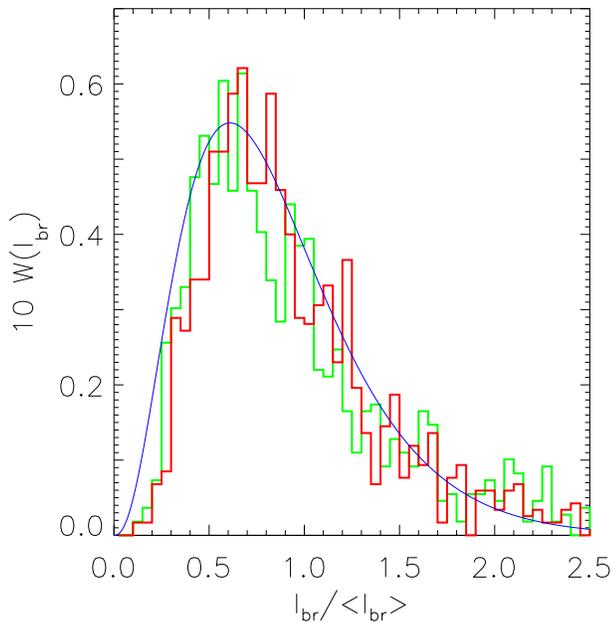}
\vspace{1.2cm}
\caption{Distribution functions, $W$, for the distance between branch 
points along a trunk for filaments selected in LDRs with linking lengths 
$r_{lnk}=3.2 ~\&~3.6 h^{-1}$Mpc (red and green lines). Fit (\ref{lbr}) 
is plotted by blue line.
} 
\end{figure}

\begin{table*}
\begin{minipage}{160mm}
\begin{center}
\caption{Wall properties in observed and simulated catalogues}
\label{tbl2}
\begin{tabular}{lcc ccc ccc} 
sample&$\langle q_w\rangle/\Gamma$&$\tau_m/\sqrt{\Gamma}$&
$\langle\delta_r\rangle$&$\langle\delta_t\rangle$&$\langle h_r\rangle$&
$\langle h_t\rangle$&$\langle w_w \rangle$&$\langle D_{sep}\rangle$\cr
  &  &  &  & &$h^{-1}$Mpc&$h^{-1}$Mpc& km/s&$h^{-1}$Mpc \cr
\hline
\multicolumn{9}{c}{radial cores}\cr 
S-380&$2.28\pm 0.58$&$0.58\pm 0.07$&1.5&-&$10.2\pm 1.6$&-&$293\pm 47$&$53\pm ~~8$\cr 
S-600&$2.26\pm 0.65$&$0.58\pm 0.08$&1.4&-&$10.5\pm 1.8$&-&$302\pm 52$&$58\pm   11$\cr 
N-380&$2.83\pm 0.73$&$0.65\pm 0.09$&1.8&-&$10.9\pm 1.4$&-&$316\pm 40$&$70\pm  ~~9$\cr 
N-600&$2.57\pm 0.76$&$0.62\pm 0.09$&1.3&-&$13.5\pm 2.8$&-&$389\pm 81$&$74\pm   12$\cr 
SDSS EDR  &$2.47\pm 0.72$&$0.61\pm 0.09$&1.5&-&$11.4\pm 2.5$&-&$329\pm 71$&$68\pm   13$\cr 
\hline
\multicolumn{9}{c}{transverse cores}\cr 
S-380&$2.39\pm 0.78$&$0.60\pm 0.10$&- &2.5& -&$6.3\pm 1.3$&   -   &$57\pm  12$\cr 
S-600&$2.29\pm 0.64$&$0.58\pm 0.08$&- &3.9&- &$4.0\pm 0.8$&   -   &$58\pm  11$\cr
N-380&$2.57\pm 0.74$&$0.62\pm 0.09$&- &3.1&- &$5.5\pm 1.0$&   -   &$77\pm  14$\cr
N-600&$2.47\pm 0.51$&$0.61\pm 0.07$&- &3.9&- &$4.3\pm 0.8$&   -   &$65\pm  11$\cr
SDSS EDR  &$2.42\pm 0.67$&$0.60\pm 0.08$&- &3.5&- &$4.9\pm 1.3$&   -   &$64\pm  14$\cr
\hline
\multicolumn{9}{c}{observed samples}\cr 
  SDSS EDR&$2.46\pm 0.7$&$0.60\pm 0.09$&1.5&3.5&$11.4\pm 2.5$&$4.9\pm 1.3$&$329\pm 71$&$66\pm 13$\cr
      LCRS&$2.51\pm 0.9$&$0.62\pm 0.10$&3.0&7.4&$~~8.6\pm 0.8$&$2.8\pm 0.7$&$247\pm 48$&$60\pm 10$\cr
      DURS&$2.23\pm 0.6$&$0.58\pm 0.08$&1.7&6.5&$~~9.7\pm 1.8$&$4.9\pm 1.2$&$280\pm 52$&$44\pm 10$\cr
\hline
\multicolumn{9}{c}{mock catalogues in real and redshift spaces for the model with $\Gamma=0.2$}\cr 
redshift&$2.7\pm 0.5$&$0.63\pm 0.06$&1.8&3.8&$11.8\pm 2.1$&$6.5\pm 1.4$&$338\pm 65$&$50\pm  10$\cr
real    &$2.1\pm 0.4$&$0.57\pm 0.06$&4.3&4.6&$~~4.8\pm 1.0$&$4.2\pm 1.0$&$305\pm 47$&$50\pm  10$\cr
\hline
\end{tabular}
\end{center}
\end{minipage}
\end{table*}

\section{Parameters of the wall--like structure elements}

The statistical characteristics of observed walls were first measured
using the LCRS and DURS (Doroshkevich et al. 2000; 2001). The rich
sample of walls extracted from the SDSS EDR, however, permits more
refined estimates of the wall properties. As was discussed in Sec. 3.2
walls dominate the HDRs, and thus these subsamples of galaxies can be
used to estimate the wall properties.

The expected characteristics of walls and methods of their measurement
were discussed in Demia\'nski et al. (2000); so here we will only
briefly reproduce the main definitions. It is important that these
characteristics can be measured independently in radial and transverse
directions, which reveals the strong influence of the velocity
dispersion on other wall characteristics.

\subsection{Main wall characteristics}

Main characteristic of walls is their mean dimensionless surface
density, $\langle q_w\rangle$, measured by the number of galaxies per
1 Mpc$^2$ and normalized by the mean density of galaxies multiplied by
a coherent length of initial velocity field (DD99; DD02)
\be
l_v\approx 33 h^{-1}{\rm Mpc}~(0.2/\Gamma),\quad \Gamma = \Omega_m h\,,
\label{lv}
\ee
where $\Omega_m$ is the mean matter density of the Universe. 
For Gaussian initial perturbations, the expected probability 
distribution function (PDF) of the surface density is 
\be
N_m(q_w) = {1\over\sqrt{2\pi}}{1\over\tau_m\sqrt{q_w}}\exp\left(
-{q_w\over 8\tau_m^2}\right){\rm erf}\left(\sqrt{q_w\over 
8\tau_m^2}\right)\,,
\label{wq}
\ee
\[
\langle q_w\rangle = 8(0.5+1/\pi)\tau_m^2\approx 6.55\tau_m^2\,.
\]
This relation links the mean surface density of walls with the 
dimensionless amplitude of perturbations, $\tau_m$,
\be
\tau_m=\sqrt{\langle q_w\rangle /6.55}\,, 
\label{taum}
\ee
which can be compared with those measured by other methods (DD02). 

Other important characteristics of walls are the mean velocity
dispersion of galaxies within walls, $\langle w_w \rangle$, the mean
separation between walls, $\langle D_{sep}\rangle$, the mean
overdensity, $\langle\delta\rangle$, and the mean thickness of walls,
$\langle h\rangle$. The mean velocity dispersion of galaxies, $\langle
w_w \rangle$, can be measured in radial direction only whereas other
wall characteristics can be measured both radially and along
transverse arcs. Comparison of the wall thickness and the overdensity,
$\langle h\rangle$ and $\langle\delta\rangle$, measured in transverse
($t$) and radial ($r$) directions, illustrates the influence of the
velocity dispersion of galaxies on the observed wall thickness.

The velocity dispersion of galaxies within a wall $w_w$ can be related
to the radial thickness of the wall by this relation (Demia\'nski et
al. 2000):
\be
h_r = \sqrt{12} H_0^{-1} w_w\,.
\label{ww_to_hr}
\ee

For a relaxed, gravitationally confined wall, the measured wall
overdensity, surface density, and the velocity dispersion are linked
by the condition of static equilibrium.  Consider a wall as a slab in
static equilibrium, and this slab has a nonhomogeneous matter
distribution across it.  We can then write the condition of static
equilibrium as follows:
\be
w_w^2 = {\pi G\mu^2\over\langle\rho\rangle \delta}\Theta_\Phi = 
{3\over 8}{\Omega_m\over\delta}(H_0l_vq_w)^2\Theta_\Phi\,,
\label{phi}
\ee
Here $\mu=\langle\rho\rangle l_v q_w$ is the mass surface density of
the wall and the factor $\Theta_\Phi\sim$ 1 describes the
nonhomogeneity of the matter distribution across the
slab. Unfortunately, for these estimates we can only use the velocity
dispersion and overdensity measured for radial and transverse
directions, respectively, and, so, the final result cannot be averaged
over the samples of walls.

\subsection{Measurement of the wall characteristics}

The characteristics of the walls can be measured with the two
parameter core--sampling approach (Doroshkevich et al. 1996) applied
to the subsample of galaxies selected within HDRs. With this method,
all galaxies of the sample are distributed within a set of radial
cores with a given angular size, $\theta_c$, or within a set of
cylindrical cores oriented along arcs of right ascension with a size
$d_c$. All galaxies are projected on the core axis and collected to a
set of one-dimensional clusters with a linking length, $l_{link}$. The
one-dimensional clusters with richnesses greater than some threshold
richness, $N_{min}$, are then used as the required sample of walls
within a sampling core.

Both the random intersection of core and walls and the nonhomogeneous 
galaxy distribution within walls lead to significant random scatter 
of measured wall characteristics. The influence of these factors 
cannot be eliminated, but it can be minimized for an optimal range 
of parameters $\theta_c$, $d_c$, $l_{link}$ and $N_{min}$. Results 
discussed below are averaged over the optimal range of these 
parameters. 

Ten samples of HDR galaxies were used for the measurement of wall
characteristics.  Four of these HDR samples we saw in Sec.~3: the
$\delta_{thr}=1$ \& $\delta_{thr}=0.75$ samples for N-380 and for
S-380.  The other six HDR samples were extracted from the N-600
and S-600 catalogs corrected for radial selection effects (Sec.~2.1);
the three different sets of threshold parameters employed yielded
HDRs containing $\sim$ 42\%, 44\%, and 50\% of all galaxies in the N-600
and S-600 catalogs.

For the radial measurements, the mean wall properties were
averaged over three radial core sizes ($\theta_c = 2^\circ,
2.25^\circ$ and $2.5^\circ$) and for six core-sampling linking lengths
($2 h^{-1}$Mpc$\leq l_{link}\leq 4.5 h^{-1}$Mpc).  For the transverse
measurements, the mean wall properties were averaged over four
core diameters ($d_c=$ 6.0, 6.5, 7.0, and 7.5$h^{-1}$Mpc) and
five core-sampling linking lengths ($2 h^{-1}$Mpc$\leq l_{link}
\leq 4. h^{-1}$Mpc).

\subsection{Measured characteristics of walls}

The mean radial and tranverse wall properties for the N-380, S-380,
N-600 \& S-600 catalogs are listed separately in Table~1.
Characteristics obtained by averaging over all ten samples are
compared with those from the DURS and LCRS and with those from mock
catalogues simulating the SDSS EDR (Cole 1998). The difference between
the mean walls surface densities measured for samples N-380 and S-380
-- $\sim$ 15-- 20\% -- reflects real variations in the wall properties
and an insufficient representativity of the two samples (i.e., cosmic
variance). However, the scatters of mean values listed in Table 1
partially include the dispersions depending on the shape of their
distribution functions. The actual scatter of mean characteristics of
walls averaged over eight samples listed in Table 1 is also $\leq$ 10
-- 12\%.

The amplitude of initial perturbations characterized by values
$\tau_m$ is similar for all eight listed SDSS EDR measurements and for
the LCRS and the DURS. It is quite consistent with estimates found for
simulations of the spatially flat $\Lambda$CDM cosmological model. The
measured PDF of the surface density of walls plotted in Fig. 7 is well
fitted to the expected expression (\ref{wq}). These results verify
that, indeed, the observed walls represent recently formed Zel'dovich
pancakes.

The difference between the wall thickness measured in the radial 
and transverse directions, $h_r$ and $h_t$, indicates that, 
along a short axis, the walls are gravitationally confined 
stationary objects. Just as with the `Finger of God' effect for 
clusters of galaxies, this difference characterizes the 
gravitational potential of compressed DM rather than the 
actual wall thickness. The same effect is seen as a difference 
between the wall overdensities measured in radial and transverse 
directions. 

The difference between the wall thicknesses is compared with 
the velocity dispersions of galaxies within the walls, $\langle 
w_w\rangle$. Clusters of galaxies with large velocity dispersions 
incorporated in walls also increase the measured velocity 
dispersion. The correlation between the wall surface density 
and the velocity dispersion confirms the relaxation of matter 
within walls. This relaxation is probably accelerated due to 
strong small scale clustering of matter within walls. 

\begin{figure}
\centering
\vspace{0.2cm}
\epsfxsize=7.cm
\epsfbox{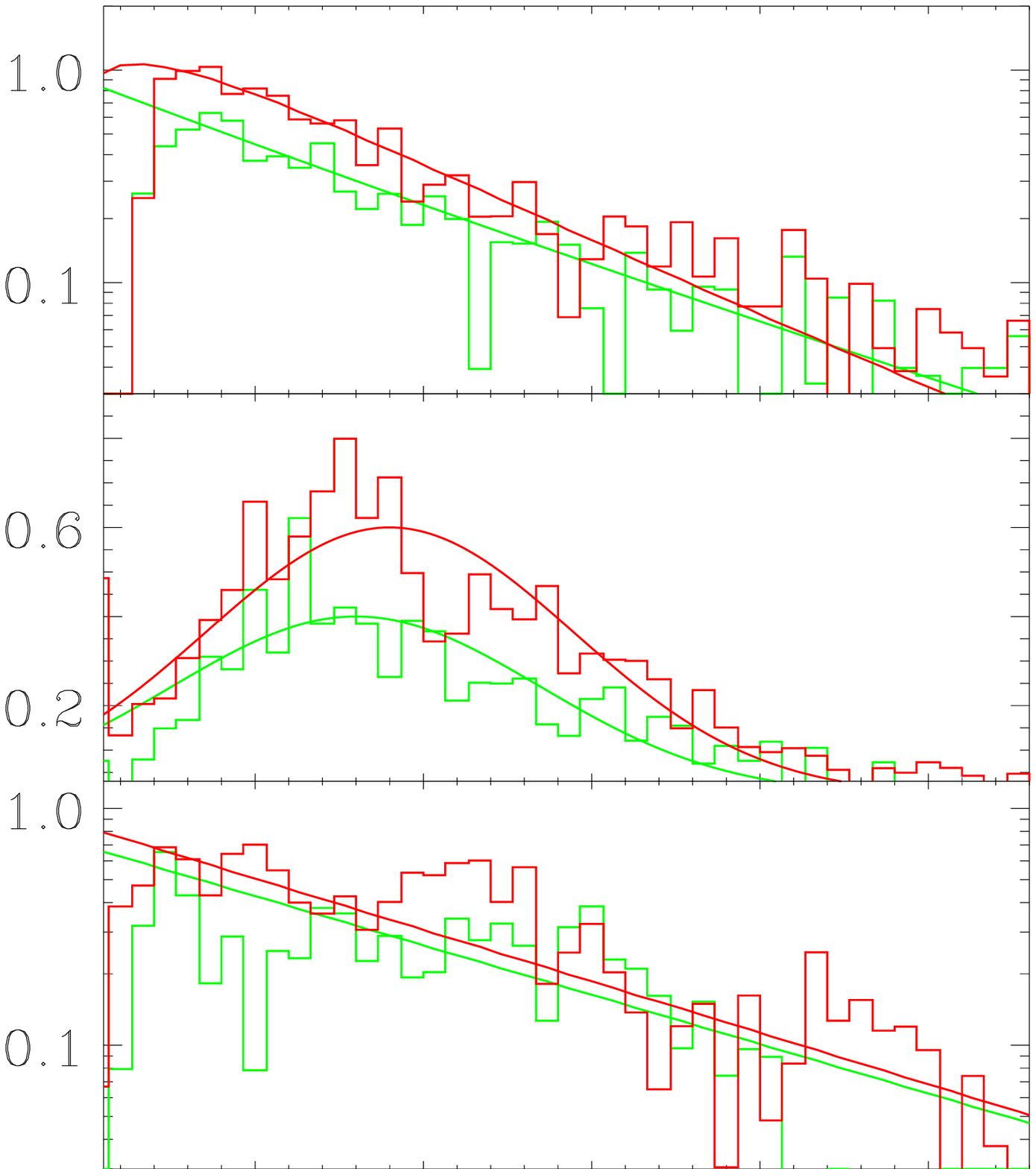}
\vspace{0.8cm}
\caption{The PDFs of observed dimensionless surface density 
of walls, $N_m(q_w/\langle q_w\rangle)$ (\ref{wq}), reduced 
velocity dispersions within walls, $N_\omega(\omega_w/\langle 
\omega_w\rangle)$ (\ref{omw}), and wall separations, $N_{sep}
(D_{sep}/\langle D_{sep}\rangle)$, averaged over radial samples 
of walls selected from samples S-380 (green lines) and N-380 
(red line). Fits (\ref{wq}) for $N_m$, Gaussian fits for 
$N_\omega$ and exponential fits for $N_{sep}$ are plotted by 
solid lines.
} 
\end{figure}

Using measured mean wall characteristics we have for the parameter 
$\Theta_\Phi$ introduced in equation~(\ref{phi})
\be
\Theta_\Phi\approx {\langle\delta\rangle\over 3}{0.3\over\Omega_m}
\approx 1\,,
\label{pphi}
\ee
which is also quite consistent with the expected value for relaxed 
and stationary walls. 

As was proposed in Demia\'nski et al. (2000) we can discriminate
between systematic variations in the measured velocity dispersion due
to regular variations in the surface density along the walls (Fig.~7,
top plot) and the random variations in the velocity dispersion due to,
for instance, random intersections of a core with rich clusters
embedded in a wall. Demia\'nski et al. (2000) suggest for
consideration a reduced velocity dispersion, $\omega_w$, \be \omega_w
= w_w q_w^{-p_w},\quad p_w\approx 0.5\,.
\label{omw}
\ee
For this reduced velocity dispersion, $\omega_w$, the systematic 
variations of $w_w$ are suppressed and the Gaussian PDF, 
$N_\omega$, can be expected. Indeed, this PDF plotted in Fig. 7 
is similar to a Gaussian function. 

Note that, for all the samples listed in Table~1, the mean wall
separation, $\langle D_{sep}\rangle$, is close to twice of the
coherent length of the initial velocity field,
\be
\langle D_{sep}\rangle\approx 2 l_v\,, 
\label{sep}
\ee
for the low density cosmological models with $\Gamma\approx 0.2$ 
(\ref{lv}). This results coincides with the estimates 
of the matter fraction -- $\sim$ 50\% -- accumulated within walls. 
Due to the large separation of walls, the correlations of their 
positions is small and a random 1D Poissonian PDF of the separation 
can be expected. These PDFs are plotted in Fig. 6 together with the 
exponential fits. 

Finally, we would like to draw attention to the fact that all measured
properties of these walls are quite consistent with a CDM--like initial
power spectrum and Gaussian distribution of perturbations.

\begin{figure}
\centering
\vspace{0.2cm}
\epsfxsize=6.7cm
\epsfbox{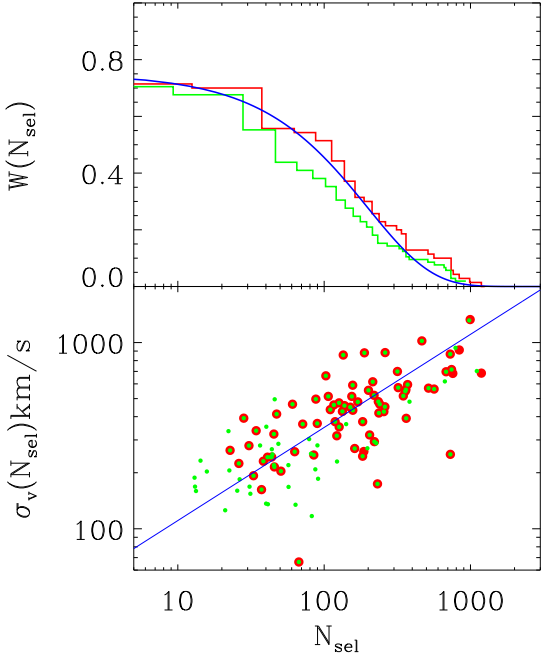}
\vspace{1.1cm}
\caption{The distribution functions, $W(N_{sel})$ (top panel), 
and velocity dispersions within possible clusters of galaxies, 
$\sigma_v(N_{sel})$ (bottom panel), are plotted vs.\ the number 
of galaxies corrected for the selection effect, $N_{sel}$, for 
two samples of clusters with $N_{mem}\geq 10$ (green lines and 
points) and with $N_{mem}\geq 15$ (red lines and points). 
Fits (\ref{wvm}) \& (\ref{vm}) are plotted by blue lines. 
} 
\end{figure}

\section{Possible rich clusters of galaxies}

The SDSS EDR also contains a number of massive galaxy clusters of
various richnesses which can be extracted by means of the MST
technique. Due to the large velocity dispersion of galaxies within
clusters and the strong `Finger of God' effect, this extraction must
be performed using different threshold linking lengths in the radial 
($r_r$) and in the transverse ($r_t$) directions.  This is not
unlike how group catalogs are extracted from redshift surveys
using conventional `friends-of-friends' algorithms (Huchra \& Geller
1982; Tucker et al. 2000).

We performed this cluster-finding in two major steps.  First, we
projected the N-600 and the S-600 samples onto a sphere of radius
$R=100 h^{-1}$ and extracted a set of candidate clusters from this 
2D galaxy catalog using a transverse linking length of $r_t=0.22
h^{-1}$Mpc {\bf ($\delta_{th}\approx$ 1)}. Second, we applied a radial
linking length of $r_r=4.5 h^{-1}$Mpc to these candidate clusters
using their (non-projected) 3D coordinates corrected for radial
selection effects (eq.~2).  In this second step, we also employed two
threshold richness, $N_{mem}=10~ \&~15$, for our final samples of
possible rich clusters.  Having extracted these probable rich
clusters, we calculated a distance-independent measure of their
richnesses by correcting their observed richnesses $N_{mem}$ for
radial selection effects using equation (1); we call this corrected
richness $N_{sel}$. Let us remind that these are {\em possible\/}
rich clusters of galaxies selected with the same algoritm; to 
confirm that they are physical potential wells, it would be best 
to check for diffuse x-ray emission.

For the threshold richness $N_{mem}=10$, 70 and 37 possible rich 
clusters with $\langle N_{sel}\rangle = 186$ were selected from the 
N-600 and S-600 samples, respectively. For the larger threshold 
richness, $N_{mem}=15$, 47 and 25 possible rich clusters with $\langle 
N_{sel}\rangle = 250$ were selected from the same samples.  The 
majority of these clusters are embedded within richer walls.  Note 
the significant excess of possible rich clusters in the north 
compared with the south -- at least for distances $D \leq 600 
h^{-1}$Mpc.

For these possible clusters, sizes in the radial and transverse 
directions differ by factors of about 40 -- 100, thus illustrating 
the `Finger of God' effect. However, this ratio is determined in 
some part by the ratio of the chosen values of the radial and 
transverse linking lengths, $r_r$ and $r_t$.  

As is seen from Fig. 8, the richness of these possible clusters 
is strongly correlated with the radial size and velocity dispersion, 
$\sigma_v$,
\be
\sigma_v\propto N_{sel}^{0.5}\,,
\label{vm}
\ee
which is quite typical for relaxed gravitationally confined
objects. For both the $N_{mem} \ge 10$ and the $N_{mem} \ge 15$
cluster samples, the distribution functions, $W(N_{sel})$, plotted in
Fig. 8 are well fitted by the exponential function
\be
W(N_{sel})\propto \exp(-N_{sel}/200 )\, .
\label{wvm}
\ee

\begin{figure*}
\begin{minipage}{160mm}
\centering
\epsfxsize=15.cm
\epsfbox{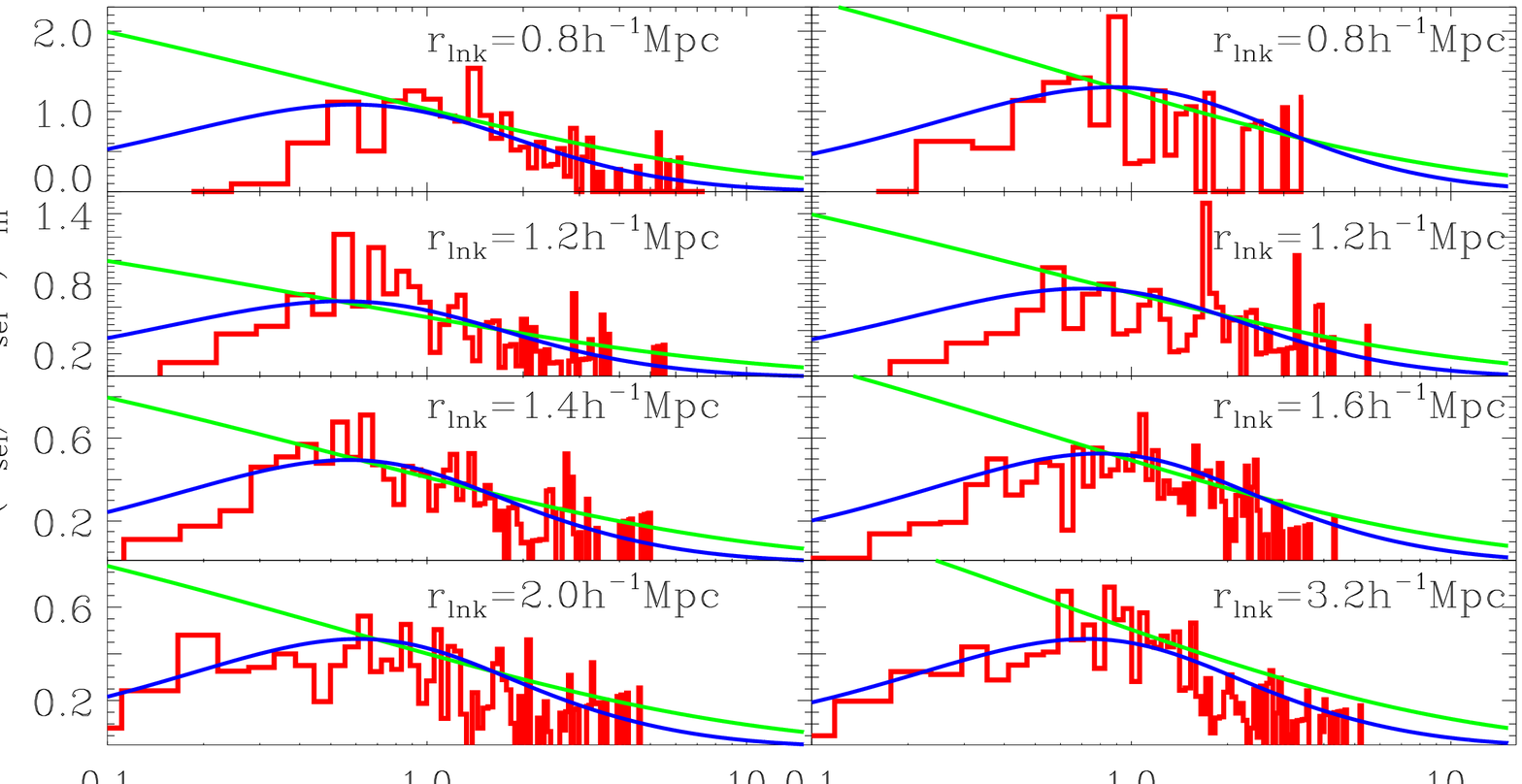}
\vspace{1.2cm}
\caption{Mass functions galaxy clouds, $N(N_{sel}/\langle
N_{sel}\rangle)$, selected in HDRs (left panels) and LDRs (right
panels) for four threshold linking lengths. Fits for relaxed
structures (eq. \ref{nm1}; $r_{lnk} = 0.8 \& 1.2 h^{-1}$~Mpc) and for
unrelaxed structures (eq. \ref{nm2}; $r_{lnk} = 1.4, 1.6, 2.0, \& 3.2
h^{-1}$~Mpc) are plotted by blue lines.  Press-Schechter fits
(eq. \ref{ps}) are plotted by green lines.  }
\end{minipage}
\end{figure*}

\section{Mass function of the structure elements}

The richness of the SDSS EDR allows one to extract several different
sets of high density clouds and structure elements with various
threshold overdensities within the HDRs and LDRs and to find their
mass function. These results can be directly compared with the
theoretical expectations of DD02.

Two samples of high density galaxy groups and two samples of
unrelaxed structure elements -- walls and filaments -- were selected
separately within HDRs and within LDRs for a threshold richness
$N_{mem}\geq$ 5. Since the velocity dispersions are expected to be
much smaller than those in rich clusters, we select these samples of
structure elements using the simpler method described in Sec.~4 rather
than the two-step approach described in Sec.~6.  The richness of each
cluster was corrected for radial selection effects using the selection
function introduced in Sec. 2.1. 

The mass functions for these samples
are plotted in Figure 9. The main properties of the selected clouds
are listed in Table~2, where $r_{lnk}$ and $\delta_{thr}$ are,
respectively, the threshold linking length and corresponding threshold
overdensity, $f_{gal}$ is the fraction of galaxies from the total
(combined HDR+LDR) sample of galaxies within the selected clouds,
$N_{cl}$ is the number of clouds, and $\langle N_{sel}\rangle$ is the
mean richness of individual clouds corrected for the selection effect.

As was shown in DD02, in Zel'dovich theory and for the WDM initial 
power spectrum the dark matter mass function of structure elements 
is independent of their shapes and, at small redshifts, it can be 
approximated by the functions
\begin{equation}
xN(x)dx = 12.5\kappa_{ZA}x^{2/3}\exp(-3.7x^{1/3})~{\rm erf}
(x^{2/3})dx\,,
\label{nm1}
\end{equation}
\begin{equation}
xN(x)dx = 8.\kappa_{ZA}x^{1/2}\exp(-3.1x^{1/3})~{\rm erf}
(x^{3/4})dx\,.
\label{nm2}
\end{equation}
\[x=\mu_{ZA}{M\over \langle M\rangle}\,,
\]
The expression (\ref{nm1}) relates to clouds which have become
essentially relaxed and static by $z=0$, and the expression
(\ref{nm2}) relates to richer, unrelaxed filaments and walls which 
are still in the process of collapse.  Here, $\kappa_{ZA}\sim$ 1.5 
-- 4 and $\mu_{ZA}\sim$ 0.8 -- 1.3
are fit parameters which take into account the incompleteness of
selected samples of clouds for small and large richnesses; this
incompleteness changes both the amplitude and mean mass of the
measured clouds. Comparison with simulations (DD02) has shown that
these relations fit reasonably well to the mass distribution of
structure elements.

\begin{table}
\caption{Parameters of groups of galaxies selected 
in HDRs and LDRs after correction for the selection effect.} 
\begin{center}
\begin{tabular}{ccc rr} 
\hline
$r_{lnk}h^{-1}$Mpc&$\delta_{thr}$&$f_{gal}$&$N_{cl}$
&$\langle N_{sel}\rangle$\\
\hline
&&HDR&&\\
0.8&45.~~&0.08&222&16.4\\
1.2&13.4 &0.2~~&403&34.2\\
1.4&~~8.3&0.25&416&44.4\\
2.0&~~2.9&0.4~~&725&90.~~\\
\hline
&&LDR&&\\
0.8& 45.~~&0.02&58&18.8\\
1.2& 13.4 &0.06&197&22.7\\
1.6&~~5.6 &0.13&396&39.6\\
3.2&~~0.8 &0.34&1176&84.8\\
\hline
\end{tabular}
\end{center}
\end{table}

For comparison, we can use the mass function from the Press-Schechter 
formalism,
\begin{equation}
xN_{PS}(x)dx = {8 \kappa_{PS}\over 45\sqrt{\pi}}~\xi^{1/6}
\exp(-\xi^{1/3})dx\,,
\label{ps}
\end{equation}
\[
\xi=1.785\mu_{ps} x=1.785\mu_{ps} M/\langle M\rangle\,.
\]
Here again the fitting parameters $\kappa_{PS}$ and $\mu_{PS}$ 
take into account the incompleteness of measured sample. This 
expression relates to the CDM-like power spectrum without small 
scale cutoff linked, for example, with the finite mass of DM 
particles, and without correction for the survival probability.
So, it does not describe the less massive part of the mass function. 

Relations (\ref{nm1}), (\ref{nm2}), \& (\ref{ps}) characterize the
mass distribution of dark matter clouds associated with the observed
galaxy groups and massive structure elements. They are closely linked 
with the initial power spectrum and quite similar
to each other despite the fact that they use different assumptions
about the process of cloud formation and the shape of the formed
clouds.  Both the Zel'dovich and Press-Schechter formalisms plotted 
in Fig. 9 fit reasonably well the observed mass distribution at
$N_{sel}\geq \langle N_{sel}\rangle$, where the exponential term in
(\ref{nm1}), (\ref{nm2}), \& (\ref{ps}) dominates.

For $N_{sel}\leq \langle N_{sel}\rangle$ the incompleteness of the 
sample of selected clouds leads to the rapid drops in the observed 
mass functions as compared with theoretical expectations. However, 
for the largest linking lengths, $r_{lnk}=2.0 ~\&~ 3.2 h^{-1}$Mpc, 
where this incompleteness is minimal, the observed mass distribution 
is quite consistent with theoretical expectations. 

For very large values of $r_{lnk}$, some excess of very massive 
clouds is generated due to the impact of the percolation process, 
which results in the formation of the network of filaments and the 
largest walls. This process is not described by the theoretical 
expressions (\ref{nm1}), (\ref{nm2})~\& (\ref{ps}). 

Results listed in Table 2 illustrate the influence of environment on
the properties of high density clouds. In particular, in spite of the
approximately equal number of galaxies in HDRs and LDRs, the majority
of the high density clouds selected with linking lengths $r_{lnk}=
0.8~\&~ 1.2 h^{-1}$Mpc are situated within the HDRs.

\section{Summary and discussion}

Statistical analysis of large galaxy redshift surveys allows us to
obtain the quantitative characteristics of large scale galaxy
distribution, which in turn can be related to the fundamental
characteristics of the Universe and the processes of structure
formation.  The large homogeneous data set compiled in the SDSS EDR
also permits us to checking the results from analysis of the LCRS and the
DURS and to obtain more accurate and more representative estimates of the
main basic characteristics of the Universe.

The spatial galaxy distribution for the N-600 and S-600 samples is
plotted in Figs. 10 \& 11;  galaxies in HDRs and in rich clusters 
are highlighted.  

\begin{figure*}
\begin{minipage}{150mm}
\centering
\epsfbox{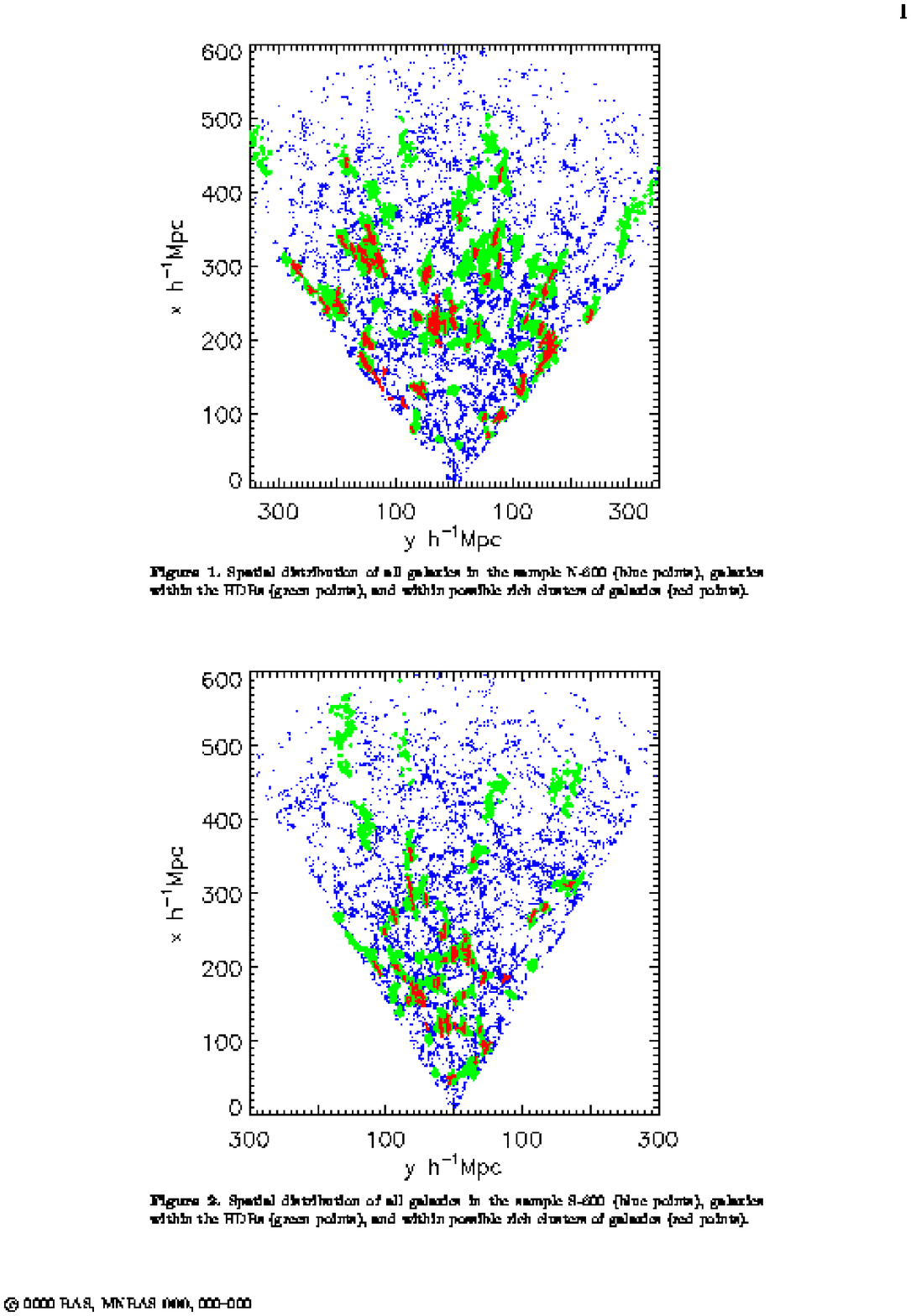}
\end{minipage}
\end{figure*}


\subsection{Main results}

The main results of our investigation can be summarized as follows:

\begin {enumerate}
\item{}
	The analysis performed in Sec. 3 with the MST technique 
	confirms that about half of galaxies are situated within 
	rich wall -- like structures and the majority of the remaining 
	galaxies are concentrated within filaments. This 
	result confirms that the filaments and walls are the main  
	structure elements in the observed galaxy distribution.
	Quantitative characteristics of walls and filaments presented 
	in Sections 3, 4 \& 5 validate this division of the LSS into these 
        two subpopulations. 
\item{}
	The typical cell size of the filamentary network is found to 
	be $\sim 10 h^{-1}$Mpc. This estimate is consistent with 
	the one obtained previously for the LCRS (Doroshkevich et al. 
	1996, 2001).
\item{}
	The main characteristics of wall--like structure elements, 
	such as the overdensity, separation distance between walls,  
	wall thickness, and the velocity dispersion within walls, 
	were measured separately for radial and transverse directions 
        in the SDSS EDR equatorial stripes.
	Comparison of these characteristics demonstrates that the 
	walls are in static equilibrium, that they are relaxed along 
	their shorter axis, and that their thickness in radial 
	direction is defined by the velocity dispersion of galaxies. 
\item{}
	The PDF of the wall surface density is consistent with 
	that predicted by Zel'dovich theory for Gaussian initial 
	perturbations. The measured amplitude of perturbations coincides 
	with that expected for a CDM-like initial power spectrum 
	and spatially flat $\Lambda$CDM cosmological model with 
	$\Omega_\Lambda\approx$ 0.7 and $\Omega_m\approx$ 0.3 .  
\item{}
	The MST technique permits the extraction of rich galaxy clusters 
	from the full observed sample of galaxies. It is found that the 
	rich selected clusters are situated mainly within richer walls 
	and that their richness correlates with the measured velocity 
	dispersion of galaxies.  
\item{}
	The mass distributions of groups of galaxies, filaments 
	and walls selected with various threshold overdensities are 
	quite well fitted to the joint mass function consistent with 
	the expectations of Zel'dovich theory.
\end{enumerate}

\subsection{Northern and southern samples}

The observations of both northern and southern samples were performed
in the same manner; so comparison of these samples can characterize
their statistical representativity.

As is seen from Figs. 1 -- 3, the general properties of both samples 
are quite similar. Thus, for both samples the radial selection 
effects are described by a single function with the same selection 
scale, $R_{sel}$.  Furthermore, the spatial galaxy distributions in 
both the N-380 and the S-380 samples are characterized by the same 
mean MST edge lengths, and the same fractions of galaxies can be 
assigned to the wall -- like and filamentary components. 

Nonetheless, the comparison of the wall properties listed in Table 
1 already shows some differences -- at the $\sim$ 10\% level -- 
between the wall richnesses for the northern and southern samples.  
A stronger difference -- by roughly a factor of 2 -- is seen in the 
number of rich galaxy clusters extracted from these two samples. 
The same north-south difference is seen in the total number of 
galaxies incorporated into these clusters.  Part of this 
factor-of-two difference, of course, is due to the different size 
areas covered by the northern and southern samples -- the northern 
sample is about $1.4\times$ larger than the southern. The remaining 
difference -- a factor of $2.0/1.4 = 1.4$ -- is likely due to cosmic 
variance.

This north-south comparison demonstrates that the achieved richness of
samples under investigation is sufficient to characterize the general
properties of the large scale spatial galaxy distribution, but it
becomes insufficient for discussing the properties of the rarer walls
and the rich clusters of galaxies.

\subsection{Properties of walls}

The walls and filaments are the largest structure elements 
observed in the Universe. In contrast to galaxies, their  
formation occurs at relatively small redshifts in course 
of the last stage of nonlinear matter clustering and is 
driven by the initial power spectrum of perturbations. 
Therefore, their properties can be successfully described 
by the nonlinear theory of gravitational instability 
(Zel'dovich 1970) that allows us to link them with the 
characteristics of initial power spectrum. 

The interpretation of the walls as Zel'dovich pancakes has been
discussed already in Thompson and Gregory (1978) and in Oort
(1983). The comparison of the statistical characteristics of the
Zel'dovich pancakes for a CDM--like initial power spectrum (DD99,
DD02) with those for observed walls demonstrates that, indeed, this
interpretation is correct and, for a given cosmological model, it
allows us to obtain independent estimates of the fundamental
characteristics of the initial power spectrum.

The estimates of the mean wall surface density, $\langle q_w
\rangle$, and the amplitude of initial perturbations, $\langle
\tau_m\rangle$, listed in Table 1 for eight samples of walls 
are consistent with each other and with  
those found for the LCRS and DURS. They are also close to those 
found for the simulated DM distribution and for the mock galaxy catalogs 
(Cole et al. 1998) prepared for a spatially flat $\Lambda$CDM 
cosmological model with $\Omega_\Lambda=0.7$ and $\Omega_m=0.3$. 

The scatter in $\langle q_w\rangle$ and $\langle\tau_m\rangle$ 
listed in Table 1 includes partly the dispersions generated by 
the shape of the distribution function of $q_w$ for a single 
measurement. Averaging of both $\langle q_w\rangle$ and $\langle
\tau_m\rangle$ listed in Table 1 for eight samples of 
walls allows us to estimate the scatter in the mean values as follows:
\be
\langle q_w\rangle = (0.49\pm 0.03)(\Gamma/0.2)\,,
\label{qmm}
\ee
\be
\tau_m = (0.27\pm 0.028)\sqrt{\Gamma/0.2}\,,
\label{staum}
\ee 
thus characterizing the variations in the mean wall properties 
for the samples under investigation. 

The amplitude of initial perturbations (eq.~\ref{staum}) is 
consistent with estimates $\tau_{CMB}$ (DD99; DD02)
\be
\tau_{CMB}\approx 0.27\sqrt{\Gamma/0.2}
\label{ttau}
\ee
obtained from the measurements of angular variations of CMB 
temperature (Bunn \& White 1997) for the same spatially flat 
$\Lambda$CDM cosmological model. The measured PDF of the 
surface density of walls plotted in Fig. 6 is well fitted 
to the expression (\ref{wq}) expected for the Gaussian initial 
perturbations (DD99; DD02). 

These results verify that, indeed, the observed walls are  
recently formed Zel'dovich pancakes. They verify also the Gaussian 
distribution of initial perturbations and coincide with the Harrison 
-- Zel'dovich primordial power spectrum. 

Comparison of other wall characteristics measured in radial 
and transverse directions indicates that the walls are 
gravitationally confined and relaxed along the shorter axis. The 
same comparison allows us to find the true wall overdensity, 
wall thickness, and the radial velocity dispersion of galaxies within 
walls. As is seen from relation (\ref{phi}), these values are 
quite self--consistent. 

\subsection{Possible rich clusters of galaxies}

Samples of possible rich clusters of galaxies extracted from the 
N-600 and S-600 samples demonstrate mainly the technical abilities 
of the MST code.  The physical reality of these clusters would 
be best tested, however, with independent x-ray observations.

Nonetheless, some characteristics of these clusters seem to be
reasonably consistent with expectations. Thus, as is seen from
Figs. 10 \& 11, these clusters are situated mainly within the richer
walls, their richness and velocity dispersion are linked by relation
(\ref{vm}), and their mass function plotted in Fig. 8 is similar to an
exponential and is dissimilar to the mass functions seen for groups of
galaxies and unrelaxed structure elements discussed in Sec. 7 . Large
differences between the number of clusters in the N-600 and S-600
samples demonstrate an insufficiently representative volume for
these relatively rare structures (remember that rich clusters typically
only contain about 10\% of all galaxies).

\subsection{Mass function of structure elements}

Rich samples of walls, filaments, and groups of galaxies in the SDSS
EDR selected using different threshold overdensities allow us to
measure their mass functions, to trace their dependence on the
threshold overdensity, and to compare them with the expectations of
Zel'dovich theory.

This comparison verifies that, for lower threshold overdensities for
both filaments and wall--like structure elements, the shape of the
observed mass functions is consistent with the expectations of
Zel'dovich theory.  For groups of galaxies, however, a deficit of low
mass groups caused presumably by selection effects and enhanced by the
restrictions inherent in our procedure for group-finding leads to a
stronger difference between the observed and expected mass functions
for $N_{sel} \leq\langle N_{sel}\rangle$.  

Let us note, that both mass functions, (\ref{nm1},\,\ref{nm2},\,\&\,
\ref{ps}), are closely linked with the initial power spectrum. They 
differ from the mass function of galaxy clusters, (\ref{wvm}), and 
the probable mass function of observed galaxies which are formed on 
account of multy step merging of less massive clouds and are described 
by the power law with an exponential cutoff (see, e.g., Silk\,\&\,
White (1978). 

\subsection{Final comments}

The SDSS (York et al. 2000) and 2dF (Colless et al. 2001) galaxy
redshift surveys provide deep and broad vistas with which cosmologists
may study the galaxy distribution on extremely large scales --
scales on which the imprint from primordial fluctuation spectrum has
not been erased.

In this paper, we have used the SDSS EDR to investigate these large
scales.  We have confirmed our earlier results, based on the LCRS and
DURS samples, that galaxies are distributed in roughly equal numbers
between two different environments: filaments, which dominate low
density regions, and walls, which dominate high density regions.
Although different in character, these two environments together form
a broken joint random network of galaxies -- the cosmic web.

Comparison with theory strongly supports the idea that the properties
of the observed walls are consistent with those for Zel'dovich
pancakes formed from a Gaussian spectrum of initial perturbations 
for a flat $\Lambda$CDM Universe ($\Omega_{\Lambda} \approx 0.7$, 
$\Omega_m\approx 0.3$). These results are well consistent to the
estimate $\Gamma=0.20\pm 0.03$ obtained in Percival et al. (2001) 
for the 2dF Galaxy Redshift Survey. 

These are important, basic conclusions regarding the large scale
structure of the Universe.  With future public releases of the SDSS
data set, we hope to refine these conclusions.

\section*{Acknowledgments}
We thank Shiyin Shen of the Max-Planck-Institut f\"ur Astrophysik 
and J\"org Retzlaff of the Max-Planck-Institut f\"ur Extraterrestrial 
Physics for useful discussions regarding this work. 

Funding for the creation and distribution of the SDSS Archive has been
provided by the Alfred P. Sloan Foundation, the Participating Institutions,
the National Aeronautics and Space Administration, the National Science
Foundation, the US Department of Energy, the Japanese Monbukagakusho,
and the Max Planck Society. The SDSS Web site is http://www.sdss.org/.
\\
The Participating Institutions are the University of Chicago, Fermilab, the
Institute for Advanced Study, the Japan Participation Group, the Johns
Hopkins University, the Max Planck Institute for Astronomy (MPIA), the
Max Planck Institute for Astrophysics (MPA), New Mexico State
University, Princeton University, the United States Naval Observatory, and
the University of Washington.
\\
This paper was supported in part by Denmark's Grundforskningsfond through 
its support for an establishment of the Theoretical Astrophysics Center.

\section*{appendix}

As mentioned in Sec.~2, we obtained our SDSS EDR sample via the SDSS
Query Tool ({\tt sdssQT}), a standalone interface to the SDSS Catalog
Archive Server.  We performed the following query to obtain our
particular sample of SDSS galaxies:

{\tt 
\noindent SELECT\\
	tag.photoobj.field.segment.run,\\
	tag.photoobj.field.segment.rerun,\\
	tag.photoobj.field.segment.camCol,\\
	tag.photoobj.field.field,\\
	tag.photoobj.objid,\\
	tag.photoobj.ra,\\
	tag.photoobj.dec,\\
	tag.g,\\
	tag.g-tag.r,\\
	z,zErr,\\
	(primTarget \& 32)+(primTarget \& 67108864)\\
FROM\\
SpecObj\\ 
WHERE\\
zConf > 0.95\\
\&\& \\
specClass == 2\\
\&\& ( (primTarget \&  64) > 0 ||\\
       (primTarget \& 128) > 0 ||\\
       (primTarget \& 256) > 0 )\\

}

This query chooses objects which were targetted as part of the main galaxy sample -- 

{\tt 
\noindent (primTarget \&  64) > 0 ||\\
          (primTarget \& 128) > 0 ||\\
          (primTarget \& 256) > 0 
}

\noindent and were found to have galaxy spectra --

\noindent {\tt specClass == 2}

\noindent As an added bonus, this query notes which of these objects
targetted as part of the main galaxy sample are also classified as
luminous red galaxies (LRGs; Eisenstein et al. 2002):

\noindent {\tt (primTarget \& 32)+(primTarget \& 67108864)}

\noindent These LRGs and their place in the general distribution of
galaxies will be discussed in a future paper (Doroshkevich, Allam, \& Tucker 2003).

\end{document}